\documentclass{ws-procs975x65}            
\begin{document}
\title{What is dimensional reduction really telling us?}

\author{Daniel Coumbe$^{a,b}$$^*$}

\address{$^a$The Niels Bohr Institute, Copenhagen University,\\
Blegdamsvej 17, DK-2100 Copenhagen, Denmark.\\
$^*$E-mail: Daniel.Coumbe@nbi.ku.dk}

\address{$^b$Faculty of Physics, Astronomy and Applied Computer Science,\\ 
Jagiellonian University, ul. prof. Stanislawa Lojasiewicza 11,\\ 
Krakow, PL 30-048, Poland\\}



\begin{abstract}

Numerous approaches to quantum gravity report a reduction in the number of spacetime dimensions at the Planck scale. However, accepting the reality of dimensional reduction also means accepting its consequences, including a variable speed of light. We provide numerical evidence for a variable speed of light in the causal dynamical triangulation (CDT) approach to quantum gravity, showing that it closely matches the superluminality implied by dimensional reduction. We argue that reconciling the appearance of dimensional reduction with a constant speed of light may require modifying our understanding of time, an idea originally proposed in Ref.~\refcite{Coumbe:2015zqa}. 

\end{abstract}


\bodymatter

\section{Introduction}\label{intro}



Given the large number of approaches to quantum gravity perhaps it is prudent to identify consistent features as a means of recognising a firm foundation on which to build. One striking example is the appearance of dimensional reduction; the idea that the number of spacetime dimensions reduces on small distances.    




Dimensional reduction was first observed in the causal dynamical triangulation (CDT) approach to quantum gravity.~\cite{Ambjorn:2005db} Since then, exact renormalisation group methods,~\cite{Lauscher:2005qz} Ho{\v r}ava-Lifshitz gravity,~\cite{Horava:2009if} loop quantum gravity,~\cite{Modesto:2008jz} and string theory, \cite{Atick:1988si} have all reported a reduction in the number of spacetime dimensions on small distance scales. The fact that dimensional reduction appears so consistently and across such a diverse number of approaches to quantum gravity strongly suggests it should be taken seriously as an indication of new Planck scale physics.   


However, if we accept the reality of dimensional reduction then logically we must also accept its consequences:

\begin{enumerate}

\item The speed of light must vary with scale.~\cite{Amelino-Camelia:2013tla,Sotiriou:2011aa}

\item Relativistic symmetries are at the very least deformed.~\cite{Amelino-Camelia:2013cfa,KowalskiGlikman:2001gp}

\item Lorentz invariance is broken in nearly all cases.~\cite{Carlip:2011tt,Sotiriou:2011mu}

\item Gravitational waves can no longer propagate.~\cite{Atick:1988si}

\item Maxwell's equations break down.~\cite{Weyl1922}

\end{enumerate}

\noindent Motivated to avoid such extreme implications, this work questions the reality of dimensional reduction, and asks whether its appearance is instead a symptom of less radical underlying Planck scale physics.



\section{Superluminality in CDT quantum gravity}

Evidence for dimensional reduction has come mainly from calculations of the spectral dimension, a measure of the effective dimension of a manifold over varying length scales. The spectral dimension $D_{S}$ is related to the probability $P_{r}$ that a random walk will return to the origin after $\sigma$ diffusion steps, and is defined by

\begin{equation}
D_{S}=-2\frac{\rm{d}\rm{log}P_{r}}{\rm{d}\rm{log}\sigma}.
\end{equation}







Independent of any particular approach to quantum gravity the spectral dimension in $(3+1)$ topological dimensions is related to the group velocity $v_{gr}$ and phase velocity $v_{ph}$ of light via


\begin{equation}
D_{S}=1+d\frac{v_{ph}}{v_{gr}} + ...\, ,
\label{GroupVel}
\end{equation}

\noindent where $d=3$ is the number of spatial topological dimensions.~\cite{Sotiriou:2011aa} For electromagnetic waves in a vacuum one expects a dimensionless speed of light parameter $c_{m}=v_{gr}/v_{ph}=1$. However, for any degree of dimensional reduction whatsoever ($D_{S}<4$) Eq.~(\ref{GroupVel}) says that the speed of light $c_{m}$ must exceed unity. Superluminality is an unavoidable consequence of dimensional reduction.  

The canonical point in the physical phase of CDT, which has an established macroscopic 4-dimensional de Sitter geometry,~\cite{Ambjorn:2008wc} has been shown to have a scale dependent spectral dimension given by the functional form

\begin{equation}
D_{S}=a-\frac{b}{c+\sigma},
\label{funcform}
\end{equation}

\noindent where $a$, $b$ and $c$ are free fit parameters,~\cite{Ambjorn:2005db} a result also found using purely analytical methods.~\cite{Giasemidis:2012qk} By substituting Eq.~(\ref{funcform}) into Eq.~(\ref{GroupVel}) we obtain a modified speed of light $c_{m}$ implied by dimensional reduction in CDT,

\begin{equation}
c_{m}=\frac{v_{gr}}{v_{ph}}=\frac{d}{a-\frac{b}{c+\sigma}-1}.
\label{cm}
\end{equation}

\begin{figure}
\centering
\includegraphics[width=0.78\linewidth]{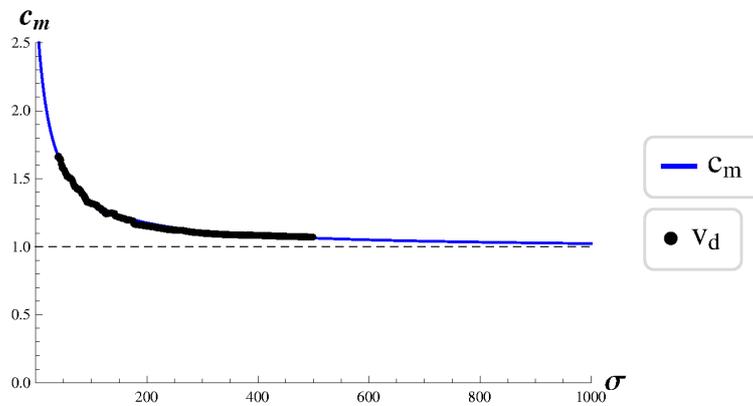}
\caption{\small The modified speed of light $c_{m}$ predicted by dimensional reduction (the solid blue curve) and the effective velocity $v_{d}$ averaged over $10^{3}$ diffusion paths in a typical CDT ensemble (filled black dots), with $C=0.18$. The dashed black line indicates the speed of light $c=1$.}
\label{SOLdiff}
\end{figure}

\noindent Figure~(\ref{SOLdiff}) shows the modified speed of light $c_{m}$ as a function of $\sigma$, with $a=4.06$, $b=135$ and $c=67$ as determined by CDT calculations.~\cite{Coumbe:2014noa}~\interfootnotelinepenalty=10000\footnote{\scriptsize An independent derivation based on dimensional reduction in CDT also gives a photon group velocity that is almost identical to $c_{m}$ when plotted as a function of $\sigma$.~\cite{Mielczarek:2015cja}}    

One can explicitly map the trajectory a fictitious diffusing particle follows in a given ensemble of triangulations defined by CDT. In CDT one approximates a continuous spacetime manifold by connecting adjacent 4-dimensional simplices via their mutual tetrahedra, forming a discretised simplicial geometry. The resulting ensemble of triangulations can be used to analyze how a test particle diffuses throughout the geometry. Starting from a randomly chosen simplex a test particle performs a random walk within the geometry by stepping between adjacent simplices $\sigma$ times. 

\begin{figure}
\centering
\includegraphics[width=0.4\linewidth]{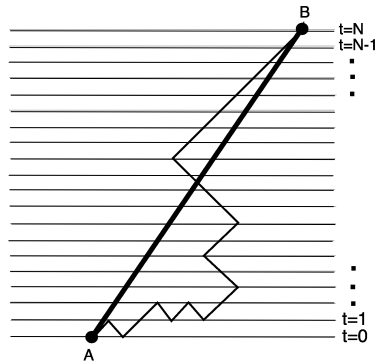}
\caption{\small A schematic representation of a test particle diffusing between points A and B in $(1+1)$-dimensional CDT.}
\label{SOLschem}
\end{figure}

A key feature of CDT is that space-like and time-like links on the lattice are distinguishable, thus allowing the foliation of spacetime into space-like hypersurfaces. In CDT, space-like hypersurfaces are separated by time intervals $t=0$, $t=1$, ..., $t=N$, as shown schematically in Fig. \ref{SOLschem}, thereby introducing a time coordinate. The elapsed time observed by a particle diffusing between points A and B is given by the number of times $t_{d}$ the test particle crosses a space-like hypersurface. This allows us to define an effective velocity $v_{d}$ for the diffusing test particle within the triangulation via $v_{d}=\frac{C\sigma}{t_{d}}$, as shown in Fig. \ref{SOLdiff}, where $C$ is a numerical constant encoding the lattice spacing.\interfootnotelinepenalty=10000\footnote{\scriptsize Since one cannot define a speed of light in Euclidean signature (in which the spectral dimension is almost universally studied) one must instead define an effective velocity $v_{d}$.}

The important feature of Fig. \ref{SOLdiff} is that the measured effective velocity of the diffusing particle $v_{d}$ closely matches the scale dependent speed of light of Eq.~(\ref{cm}), providing numerical evidence of superluminality on small distance scales in CDT.

\section{Removing superluminality}

Given the radical implications of dimensional reduction in the form of superluminal motion we explore the possibility that dimensional reduction is not a real physical phenomenon, and that its appearance in quantum gravity is instead a symptom of deeper, more conservative, Planck scale physics. Since diffusion processes that define the spectral dimension in CDT are known to diffuse within a fractal geometry, \cite{Ambjorn05} their path length should increase as a function of increasing resolution, a general property of fractal curves that has also been established for quantum mechanical paths.~\cite{AbbottWise} We contend that such a scale dependent path length is responsible for the appearence of dimensional reduction and hence superluminality in quantum gravity.  

Integration of Eq.~(\ref{funcform}) gives the probability $P_{r}$ that the diffusing particle will return to the origin after $\sigma$ diffusion steps,

\begin{equation}
P_{r}=\frac{1}{\sigma^{a/2}\left(1+\frac{c}{\sigma}\right)^{\frac{b}{2c}}}.
\end{equation}

\noindent CDT simulations find that $a\simeq 4$ and $b/2c \simeq 1$,~\cite{Ambjorn:2005db,Coumbe:2014noa} and so

\begin{equation}
P\left(\sigma\right) \simeq \frac{1}{\sigma^{2}+c \sigma}.
\label{RetProb2}
\end{equation}

The probability of return in the absence of dimensional reduction is given by $P\left(\sigma\right)=\sigma^{-2}$. Since the path length of a diffusing particle is proportional to the number of diffusion steps $\sigma$, we ask what function $\Gamma(\sigma)$ rescales the path length such that we obtain the probability of return found in CDT, namely that of Eq.~(\ref{RetProb2}). To answer this we write the equation

\begin{equation}
\frac{1}{\Gamma\left(\sigma\right)^{2}\sigma^{2}}=\frac{1}{\sigma^{2}+c \sigma},
\end{equation}

\noindent which gives

\begin{equation}
\Gamma\left(\sigma\right)=\sqrt{1+\frac{c}{\sigma}}.
\label{GammaFuncSig}
\end{equation}

\noindent In other words, the appearance of dimensional reduction and superluminality in CDT can be explained by a path length that scales according to $\Gamma\left(\sigma\right)$ as a function of $\sigma$.

Since $\sigma$ is proportional to the square of the distance scale $\Delta x$ with which one probes the manifold, and assuming the free fit parameter $c$ can be expressed as $c=Al_{p}^{2}$ in Planck units as suggested in Ref.~\refcite{Ambjorn:2005db}, where $l_{p}$ is the Planck length and $A$ is a numerical constant, then Eq.~(\ref{GammaFuncSig}) becomes

\begin{equation}
\Gamma(\Delta x)=\sqrt{1+\frac{Al_{p}^{2}}{\Delta x^2}}.
\end{equation}

Now, consider the massless diffusing particle to be a photon. In analogy with special relativity, we set up a light-clock such that each time the photon traverses the distance between two parallel mirrors defines the tick of a clock. As discussed, the photon’s path length will increase in response to a decreasing $\Delta x$ in accordance with $\Gamma(\Delta x)$. Therefore, if we are to preserve a constant speed of light in spite of such an increasing path length then the light-clock must tick slower by the same $\Gamma(\Delta x)$ factor; time must dilate as a function of relative scale to maintain an invariant speed of light.

Defining $\Delta t$ as the time it takes a photon to traverse the distance between the mirrors along the shortest possible path, and $\Delta t'$ as the time it takes when probed with resolution $\Delta x$, we obtain a relation with the same form as time dilation in special relativity, namely

\begin{equation}
\Delta t'=\Gamma(\Delta x) \Delta t.
\end{equation}

For $\Delta x \gg l_{P}$ we have $\Gamma(\Delta x) \rightarrow 1$, and so recover $\Delta t'=\Delta t$. However, for $\Delta x \approx l_{P}$ the time dilation factor $\Gamma(\Delta x)$ starts to significantly deviate from unity and may therefore modify dynamics at the Planck scale.

\section{Conclusions}

We highlight the fact that dimensional reduction has a number of radical and possibly unphysical implications, including a variable speed of light. We provide numerical evidence of superluminality in CDT quantum gravity, showing that it closely matches the variable speed of light implied by dimensional reduction. We argue that if we are to reconcile dimensional reduction and a constant speed of light then duration must be scale dependent, an idea originally proposed in Ref.~\refcite{Coumbe:2015zqa}. 



\section{Acknowledgments}

I wish to acknowledge the support of the grant DEC-2012/06/A/ST2/00389 from the National Science Centre Poland. 



\bibliographystyle{unsrt}
\bibliography{Master}


\end{document}